\documentstyle[11pt,newpasp,twoside]{article}

\def\gtrapprox{\;\lower 0.5ex\hbox{$\buildrel >\over \sim\ $}}
\def\lessapprox{\;\lower 0.5ex\hbox{$\buildrel < \over \sim\ $}}

\def\Msun  {${\rm M}_\odot$}
\def\deg   {$^\circ$}

\def\HI    {H{$\rm\scriptstyle I$}}

\def\kms   {\ km s$^{-1}$}
\def\Mhi   {M$_{\rm HI}$}

\def\edcomment#1{\iffalse\marginpar{\raggedright\sl#1\/}\else\relax\fi}
\reversemarginpar

\begin{document}
\title{The Galaxy's Eating Habits}
 \author{M. E. Putman}
\affil{Center for Astrophysics and Space Astronomy, University of Colorado, 
Boulder, CO 80309-0389; mputman@casa.colorado.edu; Hubble Fellow}
\author{C. Thom, B. K. Gibson}
\affil{Centre for Astrophysics \& Supercomputing, Swinburne University, Hawthorn, VIC, Australia 3122; cthom, bgibson@astro.swin.edu.au}
\author{L. Staveley-Smith}
\affil{Australia Telescope National Facility, CSIRO, P.O. Box 76, Epping, NSW 1710 Australia; Lister.Staveley-Smith@csiro.au}

\begin{abstract}
The possibility of a gaseous halo stream which was stripped
from the Sagittarius dwarf galaxy is presented.  
The total mass of the neutral hydrogen along the orbit of the Sgr dwarf
in the direction of the Galactic Anti-Center is $4 - 10 \times 10^6$ \Msun\ 
(at 36 kpc, the distance to the stellar debris in this region).
Both the stellar and gaseous components have negative velocities in this
part of the sky, but the gaseous component extends to higher negative velocities.
We suggest this gaseous stream was stripped from the main body of the dwarf 
0.2 - 0.3 Gyr ago during its
current orbit after a passage through a diffuse edge of the Galactic
disk with a density $> 10^{-4}$ cm$^{-3}$.  The gas would then represent the dwarf's 
last source of star formation fuel and explains how the galaxy was forming
stars 0.5-2 Gyr ago.
\end{abstract}

\section{Introduction}

Our Galaxy is in the process of consuming its small neighbors.
Though this may sound rather grim, it is its only source of fuel
and has allowed our Galaxy to grow into the large spiral it
is today.  
One of the closest
examples of a recognizable accreting satellite is the Sagittarius dwarf galaxy (28 kpc; hereafter 
Sgr dwarf; Ibata et al. 1994).
The evidence continues to accumulate that contiguous 
streams of leading and trailing stellar debris are being pulled from
the Sgr dwarf as it spirals into the Milky Way (e.g., Newberg et al. 2003), with the recent 2MASS
data tracing this trail around the majority of the Galaxy (Majewski et al. 2003; hereafter M03).  The
stars associated with the Sgr dwarf span a wide range of ages, with the youngest
population between 0.5 - 3 Gyr old (Layden \& Sarajedini 2000; Dolphin
2002; M03).  This indicates that within the past Gyr, and during
its current orbit about the Galaxy, the Sgr dwarf
was forming stars and had a source of star formation fuel.

Neutral hydrogen is a principal source of star formation
fuel for a galaxy.  Galaxies which contain HI are commonly currently 
forming stars (e.g., Lee et al. 2002) and those without detectable HI tend to have primarily an older stellar
population and thus appear to have exhausted their star formation
fuel (e.g., Gavazzi et al. 2002).  This is evident in the dwarf
galaxies of the Local Group.  The stars in the
Local Group dwarfs vary from being almost entirely ancient ($>$ 10 Gyr;
e.g., Ursa Minor) to a number of systems which are actively forming stars
(e.g., WLM, Phoenix, LMC).  
The majority of the Local Group galaxies
with HI have formed stars within the past 2 Gyr and those
with no evidence for recent star formation do not contain detectable HI (e.g., Mateo
1998; Dolphin 2002; Bouchard et al. 2003).

HI observations of the central region of the Sgr dwarf
($\alpha$, $\delta$ = 19$^{h}$ 00$^{m}$, -30\deg\ 25\arcmin\ (J2000) indicate 
that our closest satellite galaxy does not currently
contain a significant amount of star formation fuel (\Mhi\ $< 1.5
\times 10^{4}$ \Msun\ (3$\sigma$); Koribalski, Johnston \& Otrupcek
1994).  The search for HI associated with the Sgr dwarf was continued 
by Burton \& Lockman (1999); but they also found no associated gas over 18
deg$^2$ around the core of the Sgr dwarf with limits of \Mhi\ $< 7 \times
10^{3}$ \Msun\ (3$\sigma$).  These results are surprising considering
the Sgr dwarf was forming stars within the last Gyr.  Since the
orbit of the Sgr dwarf is approximately 0.7 Gyr (Ibata \& Lewis 1998), one might 
expect to find this fuel stripped
along the dwarfs orbit, possibly at a similar location to the stellar
trail.  The trailing stellar tidal tail of the Sgr dwarf has
recently been found to extend for over 150\deg\ across the South
Galactic Hemisphere with a mean distance between 20 to 40 kpc from the
Sun (M03).  Here we present HI data from HIPASS
(HI Parkes All-Sky Survey\footnote{The Parkes Telescope is part of the Australia Telescope 
which is funded by the Commonwealth of Australia for operation as a National 
Facility managed by CSIRO.}) along the entire Sgr dwarf galaxy orbit to
investigate the possibility that a gaseous Sgr trail is also present.
The gas detected represents a potential method of tracing the
history, make-up, and classification of the Sgr dwarf galaxy, as well
as the construction of our Galaxy.

\section{Observations}

The neutral hydrogen data are from the \HI\ Parkes All--Sky Survey
(HIPASS) reduced with the {\sc minmed5} method (see Putman et al. 2003a for a 
description of this reduction method).
HIPASS is a survey for \HI\ in the Southern sky, extending from the
South celestial pole to Decl. $= +25$\deg, over velocities from $-1280$
to $+12700$ km s$^{-1}$ (Barnes  et al. 2001).  The survey utilized the 64--m Parkes radio telescope,
with a focal--plane array of 13 beams arranged in a hexagonal grid, to
scan the sky in $8^\circ$ zones of Decl. with Nyquist sampling.  The
{\sc minmed5} reduced HIPASS data has a spatial resolution of
15.5\arcmin\ and a spectral resolution, after Hanning smoothing, 
of 26.4 km s$^{-1}$.   
For extended sources, the RMS noise is 10 mJy
beam$^{-1}$ (beam area 243 arcmin$^2$), corresponding to a brightness
temperature sensitivity of 8 mK. 
The northern extension of the survey, from $+2$\deg\
to $+25$\deg, was only recently completed and is presented for
the first time here.  The noise in these cubes is slightly elevated
compared to the southern data (11 mK vs. 8 mK).  This may be
due to a combination of low zenith angles during these
observations and an inability to avoid solar interference as
effectively.   
Integrated intensity maps of the high positive and
negative velocity gas (generally $|$v$_{\rm lsr}| > 80$ \kms, as long as the gas was clearly separate from
Galactic emission) were made for the 24 deg$^2$ cubes which lie along the Sgr
orbit.  The positive velocity cubes had very little emission in
them, so we concentrated on the negative velocity cubes.
The noise at the edges of the negative velocity maps were blanked within
{\sc aips} and the images were then read into {\sc idl} to create the map of 
the entire orbit shown in Figure 1. 
At 20-40 kpc the {\sc minmed5} reduced HIPASS data has a sensitivity to 
clouds of gas with \Mhi\ $> 80-320$ \Msun\ ($\Delta$v = 25 \kms; 3$\sigma$).

\section{Results}

The large scale HI map which includes all of
the high negative velocity gas along the orbit of the Sgr dwarf
is shown in Figure 1.  The gas attributed to the Magellanic Stream and
Galactic Center is labeled.  
The stream of M giants presented by M03 has quite a broad width (commonly
several degrees) along this orbit. 
The only region which has both gas and stars at negative velocities is
between $\alpha =$ 2 to 5.5 h and $\delta =$ 0 to
30\deg, or $\ell \approx 155$ to 195\deg\ and $b \approx -5$ to $-50$\deg.  
This high velocity HI gas was previously
identified as part of the anticenter complexes (ACHV and ACVHV; Wakker \& van Woerden
1991), but its
relationship to the Sgr dwarf was not previously noted.
This gas extends over the velocity range 
of -380 to -125 \kms\ (LSR), with the gas clouds orientated along 
the orbit of the Sgr dwarf ($\alpha \sim 2 - 4.5$ h) in the velocity range
-380 to -180 \kms~ (see the channel maps in Putman et al. 2003b).  
The distinct filament extending vertically across the Sgr orbit
at $\alpha \sim 5$~h is in the velocity range of
-245 to -75 \kms.  All of the gas shown in
this region would have a mass of $10^7$ \Msun\ at 36 kpc, the 
approximate distance to the stars in the Sgr stream (M03; Ibata et al. 2001).  
If only the stream of clouds orientated along the orbit
of the Sgr dwarf is considered, the HI mass is $4 \times 10^6$ \Msun.  
The gas has peak column densities on the order of $10^{20}$ cm$^{-2}$ at the
15.5\arcmin\ resolution of HIPASS and extends to the column density
limits of the data ($5\sigma \sim 3 \times 10^{18}$ cm$^{-2}$; $\Delta$v $= 25$ \kms).
The carbon stars
associated with the Sgr dwarf at this position have velocities between 
-140 to -160 \kms\ (LSR; e.g., Dinescu et al. 2002).  
The negative velocity gas clouds at ($\alpha, \delta) \sim 15.25$~h, $-19.5$\deg~  
represent Complex L and are in a region of the Sgr stellar stream that
has positive velocity stars.

\section{Discussion}

The present orbit of the Sgr
dwarf is estimated to be 0.7 Gyr (Ibata \& Lewis 1998).   Using this
orbit, the core of the Sgr dwarf was at the position of the HI complex
noted here approximately 0.2 to 0.3 Gyr ago.  It would make sense if 
the gas was part of the Sgr dwarf within that timescale considering
the age of the Sgr dwarf stellar population (0.5 - 3 Gyr; Layden \& 
Sarajedini 2000; Dolphin 2002; M03). 
These gas clouds are along a sightline relatively
close to the plane of our Galaxy, and at this position the Sgr dwarf stellar
debris is approximately 45 kpc from the Galactic Center (M03). 
This distance is significantly beyond the
typical radius quoted for our Galaxy ($\sim$26 kpc), however it
is possible that an extended ionized disk exists (e.g., Savage
et al. 2003), 
as found in other systems (Bland-Hawthorn, Freeman \& Quinn 1997).
The passage through an extended disk of our Galaxy, in addition to
the tidal forces already obviously at work as evident from the stellar
tidal stream, might have been enough to disrupt the HI in the core of
the galaxy and cause the dwarf to lose all remaining star formation
fuel.  The gas will be stripped from a galaxy if $\rho_{IGM} v^2 > \sigma^2 \rho_{gas} / 3$ (Mori \& Burkert 2000; Gunn \& Gott 1972).
We can use this equation to 
estimate the density needed at the edge of the disk to strip the gas
from the core of the Sgr dwarf.  We use a
tangential velocity of 280 \kms\ for the Sgr dwarf and a velocity
dispersion of 11.4 \kms\ (Ibata et al. 2003; Ibata \& Lewis 1998).  If the column
densities and size of the HI distribution in the core of the Sgr dwarf
were on the order of $5 \times 10^{20}$ cm$^{-2}$ (averaged over the core)
and 1 kpc, the typical $\rho_{gas}$ is 0.16 cm$^{-3}$.
An extended disk density greater than $3 \times 10^{-4}$ cm$^{-3}$ is 
then needed to strip the gas via ram pressure stripping.  Based on 
previous estimates of the density of the Galactic
halo, this density would be
easily achieved in the plane of Galaxy, 20 kpc from the currently
observed edge of the disk (e.g., Sembach et al. 2003; Maloney 2003).
Moore \& Davis (1994) proposed a similar model to create the Magellanic
Stream.

The gas that was most likely once part of the Sgr dwarf is the 
stream of clouds orientated along the orbit of the Sgr dwarf
between $\alpha = 2-4.5$h and v$_{\rm LSR} =$ -380 to -180 \kms~ (Figure 1).
At 36 kpc from the sun this HI gas has a mass of $4 \times 10^6$ \Msun.
This gas is at higher negative velocities than the carbon stars with velocity determinations
in this region, but this is not unexpected with the
addition of ram pressure forces to the tidal.
If the filament extending perpendicular to the orbit
is included as once being part of the Sgr dwarf, the total HI mass
goes up to $10^7$ \Msun.
A typical total HI mass 
for a dwarf galaxy is on the order of $10^7$ to $10^8$ \Msun\ (Grebel et al. 2003),
so this gas may represent anywhere from the majority to 10\%
of the total amount of gas once associated with the Sgr dwarf.  Using
a total Sgr dwarf mass of $5 \times 10^8$ \Msun\ (M03), this
gas represents 1-2\% of its total mass.
The Sgr dwarf most likely originally had more than $10^6 - 10^7$ \Msun\ of
neutral hydrogen associated with it, as the outer gaseous component would have 
been the first thing stripped from our
closest satellite (e.g., Yoshizawa \& Noguchi 2003).
This outer gas would have had column densities between
$10^{18} - 10^{19}$ cm$^{-2}$ and has most likely either already
dispersed or been ionized, although remnants of this
gas may be present as small HVCs along the Sgr orbit.   Since there is
currently no HI associated with the core of the Sgr dwarf (Koribalski
et al. 1994), and the dwarf has stars which are 0.5 - 3 Gyr
old, the HI gas 
presented here most likely represents the 
high column density gas from the core of the Sgr dwarf which was finally stripped
when the dwarf passed through the medium in the
extended Galactic disk.  This is 
supported by the relatively high peak column densities
of the HI gas currently ($\sim 10^{20}$ cm$^{-2}$). 
For the Sgr HI gas to survive for 0.2-0.3 Gyr (the time since
the last passage of the core Sgr dwarf), not to mention the
Magellanic Stream which is thought to be $>$ 0.5 Gyr old, the gas should either be
confined by an existing halo medium or associated with significant
amounts of dark matter.  

The association of HI gas with the Sgr
stellar stream suggests that the dwarf spheroidal classification for
the Sagittarius galaxy may need to be reconsidered.
It also suggests that slightly offset HI and stellar streams may be a
common feature of disrupted satellites in the Galactic halo.  With
the accretion of this Sgr HI stream, the Magellanic Stream (Putman et al. 2003),
Complex C (Wakker et al. 1999), and possibly other HVCs, there is ample
fuel for our Galaxy's continuing star formation and understanding the
distribution of stellar metallicities (e.g., the G-dwarf problem is not a problem). 
Determining the metallicity, distance, and ionization properties of the HI
gas presented here will aid in confirming if this HI gas was indeed stripped from
the Sgr dwarf during its current orbit.  Since we propose this is the last gas
from the core of the Sgr dwarf, we will also look for molecular gas and
dust associated with these clouds and complete simulations to determine
the detailed parameters of gaseous stripping from the Sgr dwarf.

\begin{figure}
\plotfiddle{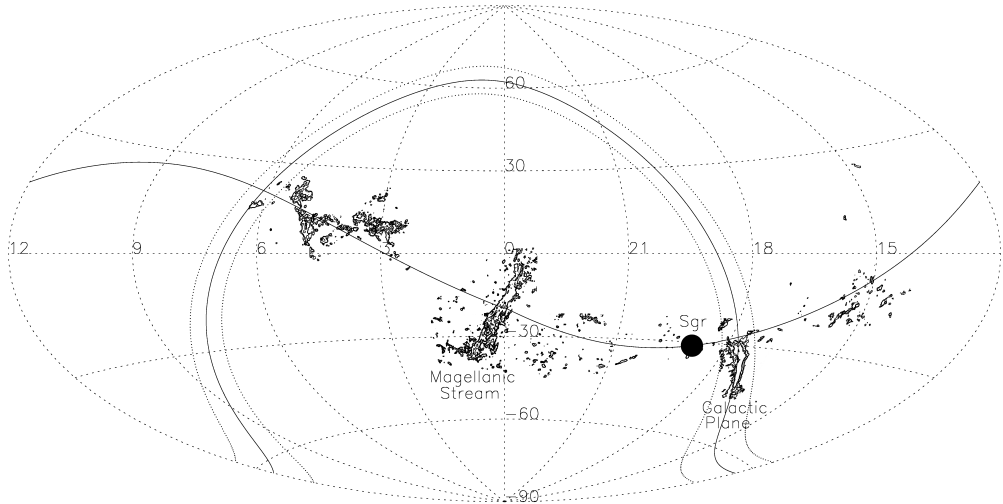}{0pt}{0}{55}{55}{-190}{-180}
\vskip 2.4in
\caption{The negative high-velocity HI ($\sim -85$ to $-400 $\kms; LSR) along the orbit of the Sgr dwarf galaxy in Celestial coordinates.  The current position of the Sgr dwarf is shown by the solid point, and the orbit of the Sgr dwarf (Ibata \& Lewis 1998) is plotted as the solid line through this point.
The negative velocity gas attributed to the Magellanic Stream and Galactic Center is labeled, and the Galactic Plane is indicated by the solid line with the two dotted lines
on each side representing $b = +5$\deg\ and $-5$\deg. The negative velocity carbon
stars extend from approximately $\alpha, \delta = 0^{h}, -20$\deg\ to $12^{h}, 20$\deg\
 along the Sgr orbit (Ibata et al. 2001).  Contours represent column density levels of 0.5, 1.0, 5.0, and 10.0 $\times 10^{19}$ cm$^{-2}$. }

\end{figure}


\acknowledgements{We would like to thank Phil Maloney, Steve
Majewski and Geraint Lewis for useful discussions. M.E.P. acknowledges support by
NASA through Hubble Fellowship grant HST-HF-01132.01 awarded by the Space Telescope Science
Institute, which is operated by AURA Inc. under NASA
contract NAS 5-26555.  B.K.G. acknowledges the support of the Australian Research
Council, through its Large Research Grant and Discovery Project schemes.}

\end{document}